\documentclass[12pt]{article}
\usepackage{amsmath}
\usepackage{amssymb}
\usepackage{amsfonts}
\usepackage{graphics}

\newcommand{\field}[1]{\mathbb{#1}}
\newcommand{\R}{\field{R}}
\newcommand{\N}{\field{N}}

\title{Deformed shape invariance symmetry and potentials in curved space with two known eigenstates}
\author{C. Quesne\thanks{Electronic mail: cquesne@ulb.ac.be}\\ 
{\small\sl Physique Nucl\'eaire Th\'eorique et Physique Math\'ematique,  Universit\'e Libre de Bruxelles,} \\ 
{\small\sl Campus de la Plaine CP229, Boulevard~du Triomphe, B-1050 Brussels, Belgium}}
\date{ }
\begin{document}
\baselineskip=22pt plus 1pt minus 1pt
%%%%%%%%%%%%%%%%%%%%%%%%%%%%%%%%%%%%%%%%%%%%%%%%%%%%%%%%%%
\maketitle
\begin{abstract}
We consider two families of extensions of the oscillator in a $d$-dimensional constant-curvature space and analyze them in a deformed supersymmetric framework, wherein the starting oscillator is known to exhibit a deformed shape invariance property. We show that the first two members of each extension family are also endowed with such a property provided some constraint conditions relating the potential parameters are satisfied, in other words they are conditionally deformed shape invariant. Since, in the second step of the construction of a partner potential hierarchy, the constraint conditions change, we impose compatibility conditions between the two sets to build potentials with known ground and first excited states. To extend such results to any members of the two families, we devise a general method wherein the first two superpotentials, the first two partner potentials, and the first two eigenstates of the starting potential are built from some generating function $W_+(r)$ (and its accompanying function $W_-(r)$).
\end{abstract}


%
%========================================================================
%
\newpage
\section{INTRODUCTION}

Since its introduction by Mathews and Lakshmanan \cite{mathews}, a classical nonlinear oscillator with a position-dependent mass and periodic solutions with an amplitude-dependent frequency has arisen a continuing interest and given rise to a lot of works concerned with its quantization, generalization to $d$ dimensions or extensions of various types.\par
%
%------------------------------------------------------------------------------------------------------
%
Its two-dimensional (and more generally $d$-dimensional) generalization is known to describe a harmonic oscillator in a space of constant curvature $\kappa = - \lambda$, where $\lambda$ is the nonlinearity parameter entering the definitions of the potential and of the position-dependent mass \cite{carinena04a}. Hence an oscillator on the sphere or in a hyperbolic space is obtained for $\kappa = - \lambda > 0$ or $\kappa = - \lambda < 0$, respectively.\par
%
%----------------------------------------------------------------------------------------
% 
The quantum version of this model has been studied and exactly solved in one \cite{carinena04b, carinena07a, schulze}, two \cite{carinena07b, carinena07c, cq15}, three \cite{carinena12}, and $d$ \cite{cq16} dimensions.\par
%
%--------------------------------------------------------------------------------------------------
%
Recently, some rational extensions of the quantum oscillator in a $d$-dimensional space of constant curvature were construted \cite{cq16}. They were shown to be exactly solvable (ES) too, with bound-state solutions written in terms of exceptional orthogonal polynomials (see, e.g., Ref.~\cite{gomez} and references quoted therein), instead of classical orthogonal polynomials for the oscillator alone. In the course of the derivation, the known equivalence of problems in curved space or with a position-dependent mass to those arising from a deformation of the canonical commutation relations \cite{cq04} was used, allowing the possibility of discussing the extended potentials in a deformed supersymmetric (DSUSY) framework \cite{bagchi} and of exploiting the deformed shape invariance (DSI) of the oscillator in curved space.\par
%
%-------------------------------------------------------------------------------------------------------------
%
More recently, other types of extensions of the quantum oscillator in a constant-curvature space were considered \cite{cq17} and shown to lead to quasi-exactly solvable (QES) Schr\"odinger equations. The latter occupy an intermediate place between ES and non-solvable ones in the sense that for some ad hoc couplings a finite number of eigenstates can be found explicitly by algebraic means, while the remaining ones remain unknown. The simplest QES problems are characterized by a hidden sl(2, $\R$) algebraic structure \cite{turbiner87, turbiner88, ushveridze, gonzalez, turbiner16} and are connected with polynomial solutions of the Heun equation \cite{ronveaux}. Generalizations of this equation are related through their polynomial solutions to more complicated QES problems. In such cases, the functional Bethe ansatz method \cite{gaudin, ho, zhang12} is very effective for deriving solutions. It is this method that was used in Ref.~\cite{cq17}, because only the simplest extensions considered there were amenable to an sl(2, $\R$) approach.\par
%
%----------------------------------------------------------------------------------------------------------
%
Apart from the sl(2, $\R$) description and the functional Bethe ansatz method, other approaches are available for constructing QES potentials. Although SUSY quantum mechanics, together with the shape invariance (SI) concept \cite{gendenshtein}, has been mostly used to construct ES potentials \cite{cooper}, its usefulness in the field of QES potentials has also been pointed out. The simplest approach consists in starting from a given QES potential with $n+1$ known eigenstates and generating a new QES one with $n$ known eigenstates by means of unbroken SUSY \cite{gango, jatkar, roy}. Two more ambitious methods aim at constructing a QES potential without the previous knowledge of another one.\par
%
%------------------------------------------------------------------------------------------
%
The first one is the conditional shape invariance (CSI) symmetry method (see \cite{chakrabarti} and references quoted therein), wherein for a given potential form depending on some parameters, a superpotential ansatz is proposed. The parameters of the latter and the ground state energy are determined in terms of the potential parameters through the SUSY Riccati equation. The number of superpotential parameters being less than that of potential ones, there appear some constraints relating both sets. If the constructed potential is translationally SI, its partner may serve as a starting potential in a second step, but the constraints turn out to be different from the first ones. In this way, a hierarchy of partner potentials can be constructed. Each step has a different set of constraint conditions and energies of at most a few levels can be obtained algebraically depending on the possibility of simultaneously satisfying such sets of constraint conditions. In practice, this can be done exactly for the first two steps, thereby generating QES potentials with known ground and first excited states. It has been recently shown, however, that this method provides a very accurate way of approximating the other unknown eigenstates \cite{bera}.\par
%
%-------------------------------------------------------------------------------------------------
% 
The second method \cite{tkachuk98} is still more ambitious because it is not restricted to SI potentials, but is based on the general form of the first two superpotentials of a hierarchy in terms of two functions. On starting from some assumption for the latter, it is possible to construct QES potentials with two known eigenstates (the ground and first excited states again). This method has also been generalized for generating QES potentials with arbitrary two known eigenstates \cite{tkachuk01}, as well as QES potentials with three known eigenstates \cite{kuliy}.\par
%
%------------------------------------------------------------------------------------------------------------
%
The aim of the present paper is to construct QES extended oscillators in a constant-curvature space with two known eigenstates. For such a purpose, we plan to start from general potentials of the kind considered in either family of extended potentials introduced in Ref.~\cite{cq17} and to adapt the methods of Refs.~\cite{chakrabarti} and \cite{tkachuk98} to a DSUSY context.\par
%
%-----------------------------------------------------------------------------------------------------------
%
In Section II, the DSUSY description of the oscillator in a constant-curvature space and its DSI property are reviewed. In Section III, the two previously introduced families of extensions of such an oscillator are considered and it is shown that the first two potentials of each of them satisfy the DSI property with constraints. In Section IV, a general method is devised in order to build QES potentials in curved space with known ground and first excited states and it is then applied to all the potentials of the two families. Finally, Section V contains the conclusion.\par
%
%================================================================
%
\section{THE OSCILLATOR IN A CONSTANT-CURVATURE SPACE AND DEFORMED SUPERSYMMETRY}

As previously shown \cite{cq16}, the Schr\"odinger equation for the $d$-dimensional oscillator in a space of constant curvature $\kappa = - \lambda$ is separable in hyperspherical coordinates and leads to the radial equation
\begin{equation}
  \left(-(1+\lambda r^2) \frac{d^2}{dr^2} - (d-1+d\lambda r^2) \frac{1}{r} \frac{d}{dr} + \frac{l(l+d-2)}{r^2}
  + {\cal V}_0(r) - {\cal E}\right) R(r) = 0,  \label{eq:SE}
\end{equation}
where $l=0$, 1, 2,~\ldots, and
\begin{equation}
  {\cal V}_0(r) = \frac{\beta(\beta+\lambda)r^2}{1+\lambda r^2} = \lambda A - \frac{\lambda A}
  {1+\lambda r^2}, \qquad A = \frac{\beta}{\lambda}\left(\frac{\beta}{\lambda}+1\right).
\end{equation}
Here, the variable $r$ runs over $(0,+\infty)$ or $(0, 1/\sqrt{|\lambda|})$ according to whether $\lambda>0$ or $\lambda<0$. The differential operator in (\ref{eq:SE}) is formally self-adjoint with respect to the measure $(1+\lambda r^2)^{-1/2} r^{d-1} dr$.\par
%
%--------------------------------------------------------------------------------------------------------
% 
Equation~(\ref{eq:SE}) can be rewritten as a deformed Schr\"odinger equation
\begin{equation}
  \left(- \sqrt{f(r)} \frac{d}{dr} f(r) \frac{d}{dr} \sqrt{f(r)} + V(r) - E\right) \psi(r) = 0,  \label{eq:SE-def}
\end{equation}
with a deformed radial momentum $\hat{\pi}_r = \sqrt{f(r)} (-{\rm i}d/dr) \sqrt{f(r)}$ and 
\begin{equation}
\begin{split}
  & f(r) = \sqrt{1+\lambda r^2}, \\
  & E = {\cal E} - \tfrac{1}{4} \lambda (d-1)^2, \\
  & \psi(r) = r^{(d-1)/2} f^{-1/2}(r) R(r),
\end{split}  \label{eq:SE-def-1}
\end{equation}
as well as
\begin{equation}
  V(r) = V_0(r) = \frac{L(L+1)}{r^2} + \lambda A - \frac{\lambda A}{f^2}, \qquad L = l + \frac{d-3}{2}.
  \label{eq:SE-def-2}
\end{equation}
Equation (\ref{eq:SE-def}) may also be considered as a position-dependent mass Schr\"odinger equation, the ordering of the mass $m(r) = 1/f^2(r)$ and the differential operator $d/dr$ being that of Mustafa and Mazharimousavi \cite{mustafa}. Bound-state wavefunctions correspond to functions $\psi(r)$ normalizable with respect to the measure $dr$, the interval of integration being $(0,+\infty)$ for $\lambda>0$ or $(0,1/\sqrt{|\lambda|})$ for $\lambda<0$.\par
%
%-----------------------------------------------------------------------------------------------------
%
Deformed Schr\"odinger equations of type (\ref{eq:SE-def}) can be discussed in terms of DSUSY \cite{bagchi}. In the simplest case of unbroken DSUSY, one introduces a rescaled potential
\begin{equation}
  V_1(r) = V(r) - E_0,  \label{eq:rescaled-V}
\end{equation}
where $E_0$ is the ground state energy of (\ref{eq:SE-def}), and one considers a pair of partner Hamiltonians
\begin{equation}
  \hat{H}_{1,2} = \hat{\pi}_r^2 + V_{1,2}(r) + E_0, \qquad V_{1,2}(r) = W^2(r) \mp f(r) \frac{dW}{dr},
  \label{eq:H-12}
\end{equation}
defined on the same interval $(0, r_{\rm max})$. Here, the superpotential $W(r)$ is related to the ground state wavefunction $\psi_0(r)$ of $\hat{H}_1$ through
\begin{equation}
  W(r) = - f \frac{d}{dr} \log \psi_0(r) - \frac{1}{2} \frac{df}{dr}
\end{equation}
or, conversely,
\begin{equation}
  \psi_0(r) \propto  f^{-1/2} \exp \left(- \int^r \frac{W(r')}{f(r')} dr'\right).  \label{eq:gs-wf}
\end{equation}
\par
%
%----------------------------------------------------------------------------------------------------
%
The two partner Hamiltonians can be written as
\begin{equation}
  \hat{H}_1 = \hat{A}^+ \hat{A}^- + E_0, \qquad \hat{H}_2 = \hat{A}^- \hat{A}^+ + E_0,
\end{equation}
in terms of a pair of first-order differential operators
\begin{equation}
  \hat{A}^{\pm} = \mp \sqrt{f(r)} \frac{d}{dr} \sqrt{f(r)} + W(r),  \label{eq:A}
\end{equation}
and they intertwine with $\hat{A}^+$ and $\hat{A}^-$ as
\begin{equation}
  \hat{A}^- \hat{H}_1 = \hat{H}_2 \hat{A}^-, \qquad \hat{A}^+ \hat{H}_2 = \hat{H}_1 \hat{A}^+.
\end{equation}
The operator $\hat{A}^-$ annihilates the ground state wavefunction $\psi_0(r)$ of $\hat{H}_1$, whereas $\hat{A}^+$ transforms the ground state wavefunction $\psi'_0(r)$ of $\hat{H}_2$ into the first excited state one $\psi_1(r)$ of $\hat{H}_1$.\par
%
%--------------------------------------------------------------------------------------------------
%
On iterating the procedure by considering $\hat{H}_2$ as the new starting Hamiltonian, one may in principle obtain another DSUSY pair of partner Hamiltonians
\begin{equation}
  \hat{H}'_{1,2} = \hat{\pi}_r^2 + V'_{1,2}(r) + E'_0, \qquad V'_{1,2}(r) = W^{\prime 2}(r) \mp f(r) 
  \frac{dW'}{dr},  \label{eq:new-H-12}
\end{equation}
where
\begin{equation}
  V'_1(r) + E'_0 = V_2(r) + E_0.  \label{eq:new-rescaled-V}
\end{equation}
From the ground state wavefunction of $\hat{H}'_1 = \hat{H}_2$, given by
\begin{equation}
  \psi'_0(r) \propto  f^{-1/2} \exp \left(- \int^r \frac{W'(r')}{f(r')} dr'\right),
\end{equation}
and corresponding to energy $E'_0$, the first excited state wavefunction of $\hat{H}_1$ with energy $E_1 = E'_0$ is then obtained through the equation
\begin{equation}
  \psi_1(r) \propto \hat{A}^+ \psi'_0(r).  \label{eq:es-wf}
\end{equation}
\par
%
%-------------------------------------------------------------------------------------------------------------------
%
In the case of the oscillator potential in curved space (\ref{eq:SE-def-2}), the construction that we have just reviewed is easy to carry out because, as shown in \cite{cq16}, to
\begin{equation}
  V_1(r) = \frac{L(L+1)}{r^2} + \frac{\beta(\beta+\lambda)r^2}{1+\lambda r^2} - E_0
\end{equation}
corresponds
\begin{equation}
  V_2(r) = \frac{(L+1)(L+2)}{r^2} + \frac{\beta(\beta-\lambda)r^2}{1+\lambda r^2} + 2\beta - E_0
\end{equation}
with $W(r)$ given by
\begin{equation}
  W(r) = - \frac{L+1}{r} f + \frac{\beta r}{f}.
\end{equation}
This means that, up to the additive constant $2\beta$, the partner in DSUSY is similar in shape and its parameters are obtained by translation, i.e., $L \to L+1$ and $\beta \to \beta - \lambda$. In other words, the oscillator in curved space is DSI, so that a whole hierarchy of Hamiltonians can be straightforwardly constructed and the starting Schr\"odinger equation is ES.\par
%
%-----------------------------------------------------------------------------------------------------
%
{}For the extensions of the oscillator potential (\ref{eq:SE-def-2}) that we are going to consider here and which are only QES, the situation is more complicated as we will proceed to show in Section~III.\par
%
%=================================================================
%
\section{CONDITIONAL DEFORMED SHAPE INVARIANCE SYMMETRY APPLIED TO EXTENSIONS OF THE OSCILLATOR IN CURVED SPACE}

\setcounter{equation}{0}

In the present Section, we plan to consider Eqs.~(\ref{eq:SE-def}) and (\ref{eq:SE-def-1}) with a potential $V(r)$ of the form
\begin{equation}
  V(r) = V^{(1)}_m(r) = \frac{L(L+1)}{r^2} + \lambda A - \frac{\lambda A}{f^2} + \lambda \sum_{k=1}^{2m} 
  B_k f^{2k},  \label{eq:fam-1}
\end{equation}
or
\begin{equation}
  V(r) = V^{(2)}_m(r) = \frac{L(L+1)}{r^2} + \lambda A - \frac{\lambda A}{f^2} - \lambda \sum_{k=1}^{2m} 
  \frac{B_k}{f^{2k+2}},  \label{eq:fam-2}
\end{equation}
where $A, B_1, B_2, \ldots, B_{2m}$ are $2m+1$ parameters and the range of $r$ is the same as in Section~II.\par
%
%xxxxxxxxxxxxxxxxxxxxxxxxxxxxxxxxxxxxxxxxxxxxxxxxxxxxxxxxxxxxxxxxxxxxxx
%
\subsection{First family of extended potentials}

Let us start with potentials (\ref{eq:fam-1}) belonging to the first family. In the present subsection, we plan to consider more specifically the cases where $\lambda>0$, $m=1$ or $m=2$, and $B_{2m}>0$.\par
%
%++++++++++++++++++++++++++++++++++++++++++++++++++++++++++++++
%
\subsubsection{First potential of the first family}

{}For $m=1$, the potential $V(r)$ of Eq.~(\ref{eq:fam-1}) depends on $L, A, B_1$, and $B_2$ (with the restriction $B_2>0$). Let us introduce a superpotential of the form
\begin{equation}
  W(r) = \frac{\xi}{r}f + \eta \frac{r}{f} + \zeta rf, \qquad \xi \le 0, \qquad \zeta>0,
\end{equation}
depending on three parameters $\xi, \eta, \zeta$. In DSUSY, the rescaled potential (\ref{eq:rescaled-V}) is represented by $V_1(r) = W^2 - f dW/dr$, as shown in (\ref{eq:H-12}). From this Riccati equation, it follows that the three unknowns $\xi, \eta$, and $\zeta$ satisfy the system of equations
\begin{equation}
\begin{split}
  & - \frac{\eta(\eta+\lambda)}{\lambda} = - \lambda A, \\
  & \frac{\eta}{\lambda}(\eta-2\zeta) + 2\xi \eta + \zeta + \lambda \xi^2 = \lambda A - E_0, \\
  & \xi(\xi+1) = L(L+1), \\
  & \zeta \left[2(\xi-1) + \frac{2\eta-\zeta}{\lambda}\right] = \lambda B_1, \\
  & \frac{\zeta^2}{\lambda} = \lambda B_2.
\end{split}
\end{equation}
The last three equations lead to the values of the unknowns
\begin{equation}
  \xi = - L -1, \qquad \zeta = \lambda \sqrt{B_2}, \qquad \frac{\eta}{\lambda} = \frac{B_1}{2\sqrt{B_2}}
  + \frac{1}{2} \sqrt{B_2} + L + 2.  \label{eq:param-W}
\end{equation}
A combination of the first two equations yields the ground state energy
\begin{equation}
  E_0 = \lambda\left(B_1 + B_2 + \frac{3B_1}{2\sqrt{B_2}} + \frac{9}{2} \sqrt{B_2} + 5\right)
  + \lambda L \left(\frac{B_1}{\sqrt{B_2}} + 3\sqrt{B_2} + 5\right) + \lambda L^2,
\end{equation}
while the first equation provides a constraint
\begin{equation}
  A = \left(\frac{B_1}{2\sqrt{B_2}} + \frac{1}{2} \sqrt{B_2} + L + 2\right)
  \left(\frac{B_1}{2\sqrt{B_2}} + \frac{1}{2} \sqrt{B_2} + L + 3\right),  \label{eq:constraint}
\end{equation}
connecting the potential parameters. To $E_0$ corresponds the ground state wavefunction (\ref{eq:gs-wf}), which can be rewritten as
\begin{equation}
  \psi_0(r) \propto r^{-\xi} f^{-\left(\frac{\eta}{\lambda}+ \frac{1}{2}\right)} e^{-\frac{1}{2} \zeta r^2},
  \label{eq:gs}
\end{equation}
where $\xi, \zeta$, and $\eta/\lambda$ are given in Eq.~(\ref{eq:param-W}).\par
%
%---------------------------------------------------------------------------------------------------------
%
The partner $V_2(r) = W^2 + f dW/dr$ of $V_1(r)$ in DSUSY (see Eq.~(\ref{eq:H-12})) can be written as
\begin{equation}
  V_2(r) = \frac{L'(L'+1)}{r^2} + \lambda A' - \frac{\lambda A'}{f^2} + \lambda B'_1 f^2 + \lambda B'_2
  f^4 + R,
\end{equation}
in terms of some new parameters $L', A', B'_1, B'_2$, and of a constant $R$, such that
\begin{equation}
\begin{split}
  &- \lambda A' = - \frac{\eta(\eta-\lambda)}{\lambda}, \\
  &\lambda A' + R = \frac{\eta}{\lambda}(\eta-2\zeta) + 2\xi\eta - \zeta + \lambda \xi^2, \\
  &L'(L'+1) = \xi(\xi-1), \\
  &\lambda B'_1 = \zeta \left[2(\xi+1) + \frac{2\eta-\zeta}{\lambda}\right], \\
  &\lambda B'_2 = \frac{\zeta^2}{\lambda}.
\end{split}
\end{equation}
On using some previous results, we get
\begin{equation}
  L' = L+1, \qquad A' = A - \frac{B_1}{\sqrt{B_2}} - \sqrt{B_2} - 2L - 4, \qquad B'_1 = B_1 + 4\sqrt{B_2},
  \qquad B'_2 = B_2,  \label{eq:param-partner}
\end{equation}
as well as
\begin{equation}
  R = - E_0 + \lambda\left(\frac{B_1}{\sqrt{B_2}} - \sqrt{B_2} + 2L + 4\right).
\end{equation}
Hence, the starting potential $V_1(r)$ is DSI, but this deformed shape invariance is not unconditionally valid since constraint (\ref{eq:constraint}) must be satisfied. We may therefore say that the potential is conditionally deformed shape invariant (CDSI).\par
%
%------------------------------------------------------------------------------------------------------------------------
% 
Let us now try to repeat the procedure and take the partner $V_2(r)$ as a starting potential $V'(r) = V_2(r)$ with ground state energy $E'_0$. Let us represent the new rescaled potential $V'_1(r) = V'(r) + E_0 - E'_0$ (see Eq.~(\ref{eq:new-rescaled-V})) as in (\ref{eq:new-H-12}) with
\begin{equation}
  W'(r) = \frac{\xi'}{r}f + \eta' \frac{r}{f} + \zeta' rf, \qquad \xi' \le 0, \qquad \zeta'>0, \label{eq:W'}
\end{equation}
where $\xi', \eta'$, and $\zeta'$ are some new parameters. By proceeding as in the first step, we obtain for the latter
\begin{equation}
  \xi' = - L -2, \qquad \zeta' = \lambda \sqrt{B_2}, \qquad \frac{\eta'}{\lambda} = \frac{\eta}{\lambda}
  +3 = \frac{B_1}{2\sqrt{B_2}} + \frac{1}{2} \sqrt{B_2} + L + 5,  \label{eq:param-W'}
\end{equation}
and for the new ground state energy
\begin{equation}
  E'_0 = \lambda\left(B_1 + B_2 + \frac{7B_1}{2\sqrt{B_2}} + \frac{21}{2} \sqrt{B_2} + 25\right)
  + \lambda L \left(\frac{B_1}{\sqrt{B_2}} + 3\sqrt{B_2} + 13\right) + \lambda L^2.
\end{equation}
There also occurs a new constraint coming from the relation $A' = \frac{\eta'}{\lambda} \left(\frac{\eta'}{\lambda}+1\right)$. On taking Eqs.~(\ref{eq:param-partner}) and (\ref{eq:param-W'}) into account, it can be written as
\begin{equation}
  A = \left(\frac{B_1}{2\sqrt{B_2}} + \frac{1}{2} \sqrt{B_2} + L + 6\right)
  \left(\frac{B_1}{2\sqrt{B_2}} + \frac{1}{2} \sqrt{B_2} + L + 7\right) - 8.  \label{eq:new-constraint} 
\end{equation}
The ground state wavefunction $\psi'_0(r)$ of the partner is given by an equation similar to (\ref{eq:gs}) with $\xi, \eta, \zeta$ replaced by $\xi', \eta', \zeta'$, respectively.\par
%
%-------------------------------------------------------------------------------------------------
%
The two constraints (\ref{eq:constraint}) and (\ref{eq:new-constraint}) are compatible provided
\begin{equation}
  B_1 = - B_2 - \sqrt{B_2}(2L+7).
\end{equation}
Then
\begin{equation}
  A = \frac{3}{4}, \qquad \frac{\eta}{\lambda} = - \frac{3}{2},
\end{equation}
and the potential
\begin{equation}
  V(r) = \frac{L(L+1}{r^2} + \frac{3}{4}\lambda - \frac{3\lambda}{4f^2} - \lambda [B_2 + \sqrt{B_2}
  (2L+7)] f^2 + \lambda B_2 f^4,  \label{eq:QES-potential}
\end{equation}
with corresponding superpotentials
\begin{equation}
  W(r) = - \frac{L+1}{r}f - \frac{3}{2}\lambda \frac{r}{f} + \lambda \sqrt{B_2}\, rf, \qquad
  W'(r) = - \frac{L+2}{r}f + \frac{3}{2}\lambda \frac{r}{f} + \lambda \sqrt{B_2}\, rf, \label{eq:W-W'}
\end{equation}
has a ground state and a first excited state whose energies are given by
\begin{equation}
  E_0 = - \lambda \left(4\sqrt{B_2} + \frac{11}{2} + 5L + L^2\right) \qquad \text{and} \qquad
  E_1 = E'_0 = \lambda \left(\frac{1}{2} - L - L^2\right),  \label{eq:E0-E1}
\end{equation}
respectively.\par
%
%--------------------------------------------------------------------------------------------------------
%
The ground state wavefunction of potential (\ref{eq:QES-potential}) is given by
\begin{equation}
  \psi_0(r) \propto r^{L+1} f e^{-\frac{1}{2}\lambda \sqrt{B_2} r^2},
\end{equation}
while, according to Eq.~(\ref{eq:es-wf}), the first excited state wavefunction can be obtained from the partner ground state
\begin{equation}
  \psi'_0(r) \propto r^{L+2} f^{-2} e^{-\frac{1}{2}\lambda \sqrt{B_2} r^2}
\end{equation}
by acting with the first-order differential operator $\hat{A}^+$, defined in (\ref{eq:A}), which can be rewritten as
\begin{equation}
  \hat{A}^+ = -f \frac{d}{dr} - \frac{L+1}{r}f - \frac{2\lambda r}{f} + \lambda \sqrt{B_2}\, rf.
\end{equation}
The result reads
\begin{equation}
  \psi_1(r) \propto r^{L+1} f^{-1} [- (2L+3) + 2\lambda \sqrt{B_2} r^2] e^{- \frac{1}{2}\lambda
  \sqrt{B_2} r^2}.
\end{equation}
Due to their behaviour for $r \to 0$ and $r \to \infty$, the functions $\psi_0(r)$ and $\psi_1(r)$ are normalizable on $(0, +\infty)$. Furthermore, $\psi_1(r)$ has a single zero on the positive half-line at $r_0 = [(2L+3)/(2\lambda\sqrt{B_2})]^{1/2}$, as it should be.\par
%
%++++++++++++++++++++++++++++++++++++++++++++++++++++++++++
%
\subsubsection{Second potential of the first family}

{}For $m=2$, the potential $V(r)$ of Eq.~(\ref{eq:fam-1}) depends on $L, A, B_1, B_2, B_3$, and $B_4$ (with the restriction $B_4>0$) and we therefore need a superpotential with an extra term
\begin{equation}
  W(r) = \frac{\xi}{r}f + \eta \frac{r}{f} + \zeta rf + \sigma rf^3, \qquad \xi \le 0, \qquad \sigma>0.
\end{equation}
From the Riccati equation satisfied by the rescaled potential (\ref{eq:rescaled-V}), we obtain the system of equations
\begin{equation}
\begin{split}
  &- \frac{\eta(\eta+\lambda)}{\lambda} = - \lambda A, \\
  &\zeta\left[2(\xi-1) + \frac{2\eta-\zeta}{\lambda}\right] - \sigma\left(\frac{2\eta}{\lambda}-3\right)
       = \lambda B_1, \\
  &\frac{\eta}{\lambda}(\eta-2\zeta) + 2\xi\eta + \zeta + \lambda\xi^2 = \lambda A - E_0, \\
  &\xi(\xi+1) = L(L+1), \\
  &\frac{\zeta^2}{\lambda} + 2\sigma\left(\xi + \frac{\eta-\zeta}{\lambda} - 2\right) = \lambda B_2, \\
  & \frac{\sigma}{\lambda}(2\zeta-\sigma) = \lambda B_3, \\
  &\frac{\sigma^2}{\lambda} = \lambda B_4.
\end{split}
\end{equation}
The last four equations determine the values of the four unknowns,
\begin{equation}
\begin{split}
  &\xi = -L-1, \qquad \sigma = \lambda \sqrt{B_4}, \qquad \zeta = \frac{\lambda}{2\sqrt{B_4}} 
       (B_3+B_4), \\
  &\frac{\eta}{\lambda} = \frac{1}{2\sqrt{B_4}}\left(B_2 + \frac{1}{2}B_3 + \frac{3}{4}B_4 -
       \frac{B_3^2}{4B_4}\right) + L + 3,
\end{split} \label{eq:param-W-bis}
\end{equation}
and a combination of the first and third relations leads to the ground state energy
\begin{align}
  E_0 &= \frac{\lambda}{2B_4} \left[\left(B_2 + \frac{1}{2}B_3 + \frac{3}{4}B_4 - \frac{B_3^2}{4B_4}
      \right)(B_3+B_4) + 16B_4\right] \nonumber\\
  &\quad {}+ \frac{\lambda}{2\sqrt{B_4}} \left(3B_2 + \frac{13}{2}B_3 + \frac{29}{4}B_4 - 
      \frac{3B_3^2}{4B_4}\right) \nonumber \\
  &\quad {}+ \lambda L \left[\frac{1}{\sqrt{B_4}}\left(B_2 + \frac{3}{2}B_3 + \frac{7}{4}B_4 -
      \frac{B_3^2}{4B_4}\right) + 7\right] + \lambda L^2.
\end{align} 
There are now two constraints coming from the first two equations
\begin{align}
  A &= \left[\frac{1}{2\sqrt{B_4}} \left(B_2 + \frac{1}{2}B_3 + \frac{3}{4}B_4 - \frac{B_3^2}{4B_4}
      \right) + L + 3\right] \nonumber \\
  &\quad {}\times \left[\frac{1}{2\sqrt{B_4}} \left(B_2 + \frac{1}{2}B_3 + \frac{3}{4}B_4 - 
      \frac{B_3^2}{4B_4}\right) + L + 4\right], \label{eq:constraint-1} \\
  B_1 &= - \frac{B_3^2}{8B_4^2}(B_3-B_4) + \frac{1}{4B_4}\left[2B_2(B_3-B_4) - \frac{3}{2} B_3B_4
      -\frac{5}{2} B_4^2\right] \nonumber \\
  &\quad {}+ \frac{1}{\sqrt{B_4}}(B_3-2B_4) - 2L\sqrt{B_4}. \label{eq:constraint-2}
\end{align}
Furthermore, the ground state wavefunction (\ref{eq:gs-wf}) reads
\begin{equation}
  \psi_0(r) \propto r^{-\xi} f^{-\left(\frac{\eta}{\lambda}+\frac{1}{2}\right)} \exp\left[- \frac{1}{2}
  (\zeta+\sigma)r^2 - \frac{1}{4} \lambda \sigma r^4\right], \label{eq:gs-bis}
\end{equation}
where the parameters are given in (\ref{eq:param-W-bis}).\par
%
%-------------------------------------------------------------------------------------------------------
% 
The partner can be written as
\begin{equation}
  V_2(r) = \frac{L'(L'+1)}{r^2} + \lambda A' - \frac{\lambda A'}{f^2} + \lambda B'_1 f^2 + \lambda B'_2
  f^4 + \lambda B'_3 f^6 + \lambda B'_4 f^8 + R,
\end{equation}
where 
\begin{equation}
\begin{split}
  &L' = L+1, \qquad A' = A - \frac{1}{\sqrt{B_4}}\left(B_2 + \frac{1}{2} B_3 + \frac{3}{4} B_4 -
       \frac{B_3^2}{4B_4}\right) - 2L - 6, \\
  &B'_1 = B_1 + \frac{2}{\sqrt{B_4}}(B_3-2B_4), \qquad B'_2 = B_2 + 8\sqrt{B_4}, \qquad B'_3 = B_3,
       \qquad B'_4 = B_4, \\
  &R = - E_0 + \lambda\left[\frac{1}{\sqrt{B_4}}\left(B_2 - \frac{1}{2} B_3 - \frac{1}{4} B_4 -
       \frac{B_3^2}{4B_4}\right) + 2L + 6\right],  
\end{split}
\end{equation}
thus showing that $V_1(r)$ is CDSI with constraints (\ref{eq:constraint-1}) and (\ref{eq:constraint-2}).\par
%
%--------------------------------------------------------------------------------------------------
%
The superpotential $W'(r)$ of Eq.~(\ref{eq:W'}) is now replaced by
\begin{equation}
  W'(r) = \frac{\xi'}{r}f + \eta' \frac{r}{f} + \zeta' rf + \sigma' rf^3, \qquad \xi' \le 0, \qquad \sigma'>0.
\end{equation}
The latter leads to the relations
\begin{equation}
\begin{split}
  \xi' &= - L - 2, \qquad \sigma' = \lambda \sqrt{B_4}, \qquad \zeta' = \frac{\lambda}{2\sqrt{B_4}}
      (B_3+B_4), \\
  \frac{\eta'}{\lambda} &= \frac{\eta}{\lambda} + 5 = \frac{1}{2\sqrt{B_4}}\left(B_2 + \frac{1}{2}B_3
      + \frac{3}{4}B_4 - \frac{B_3^2}{4B_4}\right) + L + 8, \\
  E'_0 &= \frac{\lambda}{2B_4}\left[\left(B_2 + \frac{1}{2}B_3 + \frac{3}{4}B_4 - \frac{B_3^2}{4B_4}\right)
      (B_3+B_4) + 80B_4\right] \\
  &\quad {}+ \frac{\lambda}{2\sqrt{B_4}}\left(7B_2 + \frac{33}{2}B_3 + \frac{73}{4}B_4 - \frac{7B_3^2}
      {4B_4} + 4\sqrt{B_4}\right) \\
  &\quad {} + \lambda L\left[\frac{1}{\sqrt{B_4}}\left(B_2 + \frac{3}{2}B_3 + \frac{7}{4}B_4 -
      \frac{B_3^2}{4B_4}\right) + 19\right] + \lambda L^2,
\end{split}
\end{equation}
and to the two new constraints
\begin{align}
  A &= \left[\frac{1}{2\sqrt{B_4}} \left(B_2 + \frac{1}{2}B_3 + \frac{3}{4}B_4 - \frac{B_3^2}{4B_4}
      \right) + L + 9\right] \nonumber \\
  &\quad {}\times \left[\frac{1}{2\sqrt{B_4}} \left(B_2 + \frac{1}{2}B_3 + \frac{3}{4}B_4 - 
      \frac{B_3^2}{4B_4}\right) + L + 10\right] - 12, \label{eq:new-constraint-1} \\
  B_1 &= - \frac{B_3^2}{8B_4^2}(B_3-B_4) + \frac{1}{4B_4}\left[2B_2(B_3-B_4) - \frac{3}{2} B_3B_4
      -\frac{5}{2} B_4^2\right] \nonumber \\
  &\quad {}+ \frac{1}{\sqrt{B_4}}(3B_3-4B_4) - 2L\sqrt{B_4}. \label{eq:new-constraint-2}
\end{align}
The partner wavefunction $\psi'_0(r)$ is similar to (\ref{eq:gs-bis}) with all parameters replaced by primed ones.\par
%
%----------------------------------------------------------------------------------------------------------
%
The two sets of constraints (\ref{eq:constraint-1}), (\ref{eq:constraint-2}) and (\ref{eq:new-constraint-1}), (\ref{eq:new-constraint-2}) are compatible provided
\begin{equation}
  B_3 = B_4, \qquad B_2 = - B_4 - \sqrt{B_4}\,(2L+11).
\end{equation}
We then also obtain
\begin{equation}
  A = \frac{15}{4}, \qquad B_1 = - B_4 - \sqrt{B_4}\, (2L+1), \qquad \frac{\eta}{\lambda} = - \frac{5}{2},
  \qquad \zeta = \lambda \sqrt{B_4}.
\end{equation}
Hence, the potential
\begin{align}
  V(r) &= \frac{L(L+1)}{r^2} + \frac{15}{4}\lambda - \frac{15\lambda}{4f^2} - \lambda [B_4 + \sqrt{B_4}
       \,(2L+1)] f^2  \nonumber \\
  &\quad {} - \lambda [B_4 + \sqrt{B_4}\,(2L+11)] f^4 + \lambda B_4 f^6 + \lambda B_4 f^8,
       \label{eq:QES-potential-bis}
\end{align}
with corresponding superpotentials
\begin{equation}
\begin{split}
  &W(r) = - \frac{L+1}{r}f - \frac{5}{2}\lambda \frac{r}{f} + \lambda \sqrt{B_4} rf(1 + f^2), \\
  &W'(r) = - \frac{L+2}{r}f + \frac{5}{2}\lambda \frac{r}{f} + \lambda \sqrt{B_4} rf(1 + f^2),
\end{split} \label{eq:W-W'-bis}
\end{equation}
has a ground state and a first excited state whose energies are given by
\begin{equation}
  E_0 = - \lambda\left(6\sqrt{B_4} + \frac{17}{2} + 7L + L^2\right) \quad \text{and} \quad
  E_1 = E'_0 = \lambda\left(2\sqrt{B_4} + \frac{7}{2} + L - L^2\right),  \label{eq:E0-E1-bis} 
\end{equation}
respectively. The corresponding normalizable wavefunctions read
\begin{align}
  \psi_0(r) &\propto r^{L+1} f^2 \exp\left(- \lambda \sqrt{B_4}\,r^2 - \frac{\lambda^2}{4} \sqrt{B_4}\,
      r^4\right), \nonumber \\
  \psi_1(r) &\propto r^{L+1} f^{-2} [- (2L+3) + 4\lambda\sqrt{B_4}\,r^2 + 2\lambda^2\sqrt{B_4}\,
      r^4] \\
  &\quad {}\times \exp\left(- \lambda \sqrt{B_4}\,r^2 - \frac{\lambda^2}{4} \sqrt{B_4}\,r^4\right).
      \nonumber
\end{align}
The single zero of $\psi_1(r)$ on the positive half-line is now at $r_0 = \frac{1}{\sqrt{\lambda}} \{[(2L+3 + 2\sqrt{B_4})/(2\sqrt{B_4})]^{1/2} - 1\}^{1/2}$.\par
%
%---------------------------------------------------------------------------------------------------------
%
In Fig.~1, some examples of extended potentials (\ref{eq:QES-potential}) and (\ref{eq:QES-potential-bis}) are plotted. The corresponding wavefunctions $\psi_0(r)$ and $\psi_1(r)$ of the former are displayed in Fig.~2. Those of the latter have a similar behaviour.\par
%
%---------------------------------------------------------------------------------------------------------
%
\begin{figure}[h]
\begin{center}
\includegraphics{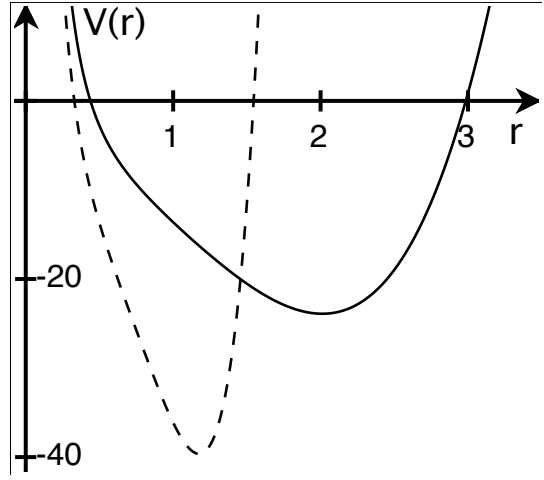}
\caption{Plots of potential (\ref{eq:QES-potential}) with $\lambda = L = B_2 = 1$ (solid line) and of potential (\ref{eq:QES-potential-bis}) with $\lambda = L = B_4 = 1$ (dashed line). The ground and first excited state energies are $E_0 = -31/2$, $E_1 = -3/2$ for the former, and $E_0 = - 45/2$, $E_1 = 11/2$ for the latter.}
\end{center}
\end{figure}
\par
%
%-------------------------------------------------------------------------------------------------------------------------------
%
\begin{figure}[h]
\begin{center}
\includegraphics{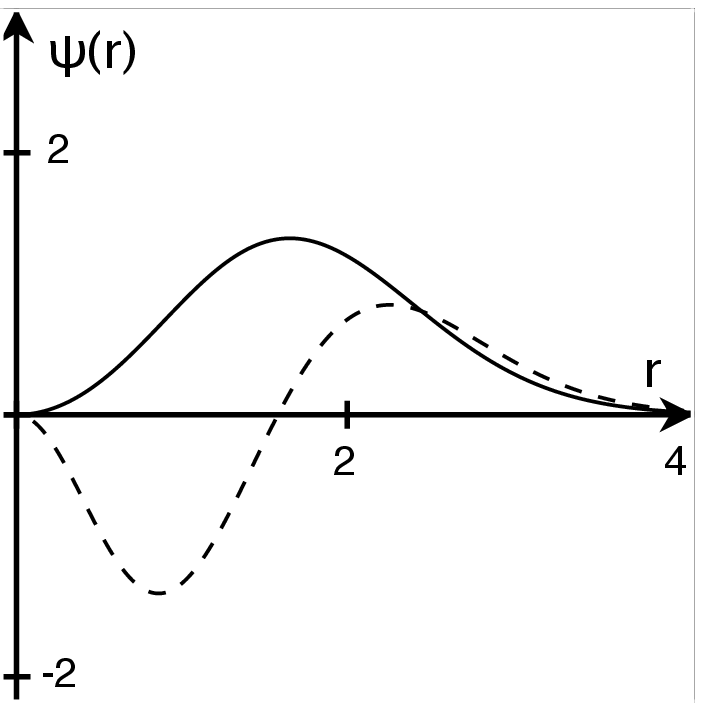}
\caption{Plots of ground state wavefunction $\psi_0(r)$ (solid line) and of first excited state wavefunction $\psi_1(r)$ (dashed line) for potential (\ref{eq:QES-potential}) with $\lambda = L = B_2 = 1$.}
\end{center}
\end{figure}
\par
%
%xxxxxxxxxxxxxxxxxxxxxxxxxxxxxxxxxxxxxxxxxxxxxxxxxxxxxxxxxxxxxxxxxx
%
\subsection{Second family of extended potentials}

Let us now consider potentials (\ref{eq:fam-2}). In the present subsection, we plan to deal more specifically with the cases where $\lambda<0$, $m=1$ or $m=2$, and $B_{2m}>0$. Since the procedure used is the same as in Section IIIA, we are not going to detail the calculations, but instead state the results.\par
%
%+++++++++++++++++++++++++++++++++++++++++++++++++++++++
%
\subsubsection{First potential of the second family}

{}For $m=1$, we consider a superpotential of the type
\begin{equation}
  W(r) = \frac{\xi}{r}f + \eta\frac{r}{f} + \zeta\frac{r}{f^3}, \qquad \xi \le 0, \qquad \zeta>0,
  \label{eq:superpot}
\end{equation}
and we obtain
\begin{equation}
\begin{split}
  &\xi = - L - 1, \qquad \zeta = |\lambda|\sqrt{B_2}, \qquad \frac{\eta}{|\lambda|} = \frac{B_1}
       {2\sqrt{B_2}} + \frac{1}{2}\sqrt{B_2} + \frac{3}{2}, \\
  &E_0 = |\lambda|\left(B_1 + B_2 + \frac{3B_1}{2\sqrt{B_2}} + \frac{9}{2}\sqrt{B_2} + \frac{11}{2}
       \right) + |\lambda| L\left(\frac{B_1}{\sqrt{B_2}} + 3\sqrt{B_2} + 5\right) + |\lambda| L^2,
\end{split}
\end{equation}
with a single constraint
\begin{equation}
  A = \left(\frac{B_1}{2\sqrt{B_2}} + \frac{1}{2}\sqrt{B_2} + \frac{3}{2}\right)
  \left(\frac{B_1}{2\sqrt{B_2}} - \frac{3}{2}\sqrt{B_2} + \frac{1}{2}\right) - 2L \sqrt{B_2}.
  \label{eq:constraint-bis}
\end{equation}
\par
%
%--------------------------------------------------------------------------------------------------------
%
The parameters characterizing the partner are given by
\begin{equation}
\begin{split}
  &L' = L+1, \qquad A' = A + \frac{B_1}{\sqrt{B_2}} - 3\sqrt{B_2} + 3, \qquad B'_1 = B_1 + 6\sqrt{B_2},
        \qquad B'_2 = B_2, \\
  &R = - E_0 + |\lambda|\left(\frac{B_1}{\sqrt{B_2}} - 3\sqrt{B_2} + 3\right),
\end{split}
\end{equation}
so that $V_1(r)$ is CDSI again.\par
%
%-------------------------------------------------------------------------------------------------------
%
The second step corresponds to a superpotential $W'(r)$ with parameters
\begin{equation}
  \xi' = - L - 2, \qquad \zeta' = |\lambda|\sqrt{B_2}, \qquad \frac{\eta'}{|\lambda|} = 
  \frac{\eta}{|\lambda|} + 3 = \frac{B_1}{2\sqrt{B_2}} + \frac{1}{2}\sqrt{B_2} + \frac{9}{2}.
\end{equation}
The new ground state energy and the new constraint are
\begin{equation}
  E'_0 = |\lambda| \left(B_1 + B_2 + \frac{7B_1}{2\sqrt{B_2}} + \frac{21}{2}\sqrt{B_2} + \frac{59}{2}
  \right) + |\lambda| L\left(\frac{B_1}{\sqrt{B_2}} + 3\sqrt{B_2} + 13\right) + |\lambda| L^2
\end{equation}
and
\begin{equation}
  A = \left(\frac{B_1}{2\sqrt{B_2}} + \frac{1}{2}\sqrt{B_2} + \frac{9}{2}\right)
  \left(\frac{B_1}{2\sqrt{B_2}} - \frac{3}{2}\sqrt{B_2} + \frac{7}{2}\right) - \frac{B_1}{\sqrt{B_2}}
  + \sqrt{B_2} - 3 - 2L \sqrt{B_2},  \label{eq:new-constraint-bis}
\end{equation}
respectively.\par
%
%---------------------------------------------------------------------------------------------------
%
The constraints (\ref{eq:constraint-bis} and (\ref{eq:new-constraint-bis}) are compatible provided
\begin{equation}
  B_1 = B_2 - 6\sqrt{B_2},
\end{equation}
thus leading to
\begin{equation}
  A = - B_2 - (2L+1)\sqrt{B_2} + \frac{15}{4}, \qquad \frac{\eta}{|\lambda|} = \sqrt{B_2} - \frac{3}{2}.
\end{equation}
We conclude that the potential
\begin{align}
  V(r) &= \frac{L(L+1)}{r^2} + |\lambda| \left[B_2 + (2L+1)\sqrt{B_2} - \frac{15}{4}\right]
       - |\lambda| \frac{B_2 + (2L+1)\sqrt{B_2} - \frac{15}{4}}{f^2} \nonumber \\
  &\quad {} + |\lambda| \frac{B_2 - 6\sqrt{B_2}}{f^4} + |\lambda| \frac{B_2}{f^6},
       \label{eq:QES-potential-ter}
\end{align}
with corresponding superpotentials
\begin{equation}
\begin{split}
  &W(r) = - \frac{L+1}{r}f + |\lambda|\left(\sqrt{B_2} - \frac{3}{2}\right) \frac{r}{f} + |\lambda|
       \sqrt{B_2} \frac{r}{f^3}, \\
  &W'(r) = - \frac{L+2}{r}f + |\lambda|\left(\sqrt{B_2} + \frac{3}{2}\right) \frac{r}{f} + |\lambda|
       \sqrt{B_2} \frac{r}{f^3},
\end{split}  \label{eq:W-W'-ter}
\end{equation}
has a ground state and a first excited state with respective energies
\begin{equation}
\begin{split}
  &E_0 = |\lambda| \left[2B_2 - \frac{7}{2} + L(4\sqrt{B_2} - 1) + L^2\right], \\
  &E_1 = E'_0 = |\lambda| \left[2B_2 + 8\sqrt{B_2} + \frac{17}{2} + L(4\sqrt{B_2} + 7) + L^2\right],
\end{split}
\end{equation}
and wavefunctions
\begin{align}
  \psi_0(r) &\propto r^{L+1} f^{\sqrt{B_2} - 2} \exp\left(- \frac{1}{2}\frac{\sqrt{B_2}}{f^2}\right), 
        \nonumber\\
  \psi_1(r) &\propto r^{L+1} f^{\sqrt{B_2} - 2} [2L+3 - 2|\lambda|(2L+3 + 2\sqrt{B_2})r^2 +
        |\lambda|^2(2L+3 + 2\sqrt{B_2})r^4] \\
  &\quad {}\times \exp\left(- \frac{1}{2}\frac{\sqrt{B_2}}{f^2}\right). \nonumber 
\end{align}
The behaviour of these functions for $r\to 0$ and $r\to 1/\sqrt{|\lambda|}$ shows that they are normalizable on $\left(0, 1/\sqrt{|\lambda|}\right)$ with respect to the measure $dr$. The single zero of $\psi_1(r)$ in this interval is located at $r_0 = \frac{1}{\sqrt{|\lambda|}} \{1 - [2\sqrt{B_2}/(2L+3+2\sqrt{B_2})]^{1/2}\}^{1/2}$.\par
%
%+++++++++++++++++++++++++++++++++++++++++++++++++++++++++++
%
\subsubsection{Second potential of the second family}

{}For $m=2$, the superpotential (\ref{eq:superpot}) is replaced by
\begin{equation}
  W(r) = \frac{\xi}{r}f + \eta\frac{r}{f} + \zeta\frac{r}{f^3} + \sigma\frac{r}{f^5}, \qquad \xi \le 0, \qquad 
  \sigma>0,
\end{equation}
and we get
\begin{align}
  \xi &= - L - 1, \qquad \sigma = |\lambda|\sqrt{B_4}, \qquad \zeta = \frac{|\lambda|}{2\sqrt{B_4}}
       (B_3 + B_4), \nonumber \\
  \frac{\eta}{|\lambda|} &= \frac{1}{2\sqrt{B_4}} \left(B_2 + \frac{1}{2}B_3 + \frac{3}{4}B_4 -
        \frac{B_3^2}{4B_4}\right) + \frac{5}{2}, \nonumber \\
  E_0 &= \frac{|\lambda|}{2B_4} \left[\left(B_2 + \frac{1}{2}B_3 + \frac{3}{4}B_4 - \frac{B_3^2}{4B_4}
       \right)(B_3+B_4) + 17B_4\right] \\
  &\quad {}+ \frac{|\lambda|}{2\sqrt{B_4}} \left(3B_2 + \frac{13}{2}B_3 + \frac{29}{4}B_4 -
       \frac{3B_3^2}{4B_4}\right) \nonumber \\
  &\quad {}+ |\lambda| L \left[\frac{1}{\sqrt{B_4}}\left(B_2 + \frac{3}{2}B_3 + \frac{7}{4}B_4 -
       \frac{B_3^2}{4B_4}\right) + 7\right] + |\lambda| L^2, \nonumber
\end{align}
together with the two constraints
\begin{align}
  A &= \left[\frac{1}{2\sqrt{B_4}} \left(B_2 + \frac{1}{2}B_3 + \frac{3}{4}B_4 -
        \frac{B_3^2}{4B_4}\right) + \frac{5}{2}\right] \nonumber \\
  &\quad {}\times \left[\frac{1}{2\sqrt{B_4}} \left(B_2 - \frac{3}{2}B_3 - \frac{5}{4}B_4 -
        \frac{B_3^2}{4B_4}\right) + \frac{3}{2}\right] - \frac{L}{\sqrt{B_4}}(B_3+B_4), \\
  B_1 &= - \frac{B_3^2}{8B_4^2}(B_3-B_4) + \frac{1}{4B_4} \left[2B_2(B_3-B_4) - \frac{3}{2}B_3B_4
       - \frac{5}{2}B_4^2\right] \nonumber \\
  &\quad {}+ \frac{1}{\sqrt{B_4}}(B_3-2B_4) - 2L\sqrt{B_4}.
\end{align}
\par
%
%---------------------------------------------------------------------------------------------------------
%
{}For the partner, we obtain the parameters
\begin{align}
  L' &= L+1, \qquad A' = A + \frac{1}{\sqrt{B_4}} \left(B_2 - \frac{3}{2}B_3 - \frac{5}{4}B_4 -
       \frac{B_3^2}{4B_4}\right) + 5, \nonumber \\
  B'_1&= B_1 + \frac{1}{\sqrt{B_4}}(3B_3-5B_4), \qquad B'_2 = B_2 + 10\sqrt{B_4}, \qquad B'_3 = B_3,
       \qquad B'_4 = B_4, \\
  R &= - E_0 + |\lambda| \left[\frac{1}{\sqrt{B_4}} \left(B_2 - \frac{3}{2}B_3 - \frac{5}{4}B_4 -
       \frac{B_3^2}{4B_4}\right) + 5\right], \nonumber
\end{align}
showing that $V_1(r)$ is CDSI as in the previous cases.\par
%
%----------------------------------------------------------------------------------------------------
%
{}For the superpotential $W'(r)$ corresponding to the second step, we get the parameters
\begin{equation}
\begin{split}
  &\xi' = - L - 2, \qquad \sigma' = |\lambda|\sqrt{B_4}, \qquad \zeta' = \frac{|\lambda|}{2\sqrt{B_4}}
       (B_3+B_4), \\
  &\frac{\eta'}{|\lambda|} = \frac{\eta}{|\lambda|} + 5 = \frac{1}{2\sqrt{B_4}} \left(B_2 + \frac{1}{2}B_3 +
        \frac{3}{4}B_4 - \frac{B_3^2}{4B_4}\right) + \frac{15}{2}.
\end{split}
\end{equation}
We have now a new ground state energy
\begin{align}
  E'_0 &= \frac{|\lambda|}{2B_4} \left[\left(B_2 + \frac{1}{2}B_3 + \frac{3}{4}B_4 - \frac{B_3^2}{4B_4}
       \right)(B_3+B_4) + 75 B_4\right] \nonumber \\
  &\quad {}+ \frac{|\lambda|}{2\sqrt{B_4}} \left(7B_2 + \frac{33}{2}B_3 + \frac{73}{4}B_4 -
       \frac{7B_3^2}{4B_4} + 18\sqrt{B_4}\right) \nonumber \\
  &\quad {}+ \frac{|\lambda|L}{\sqrt{B_4}} \left(B_2 + \frac{3}{2}B_3 + \frac{7}{4}B_4 - \frac{B_3^2}
       {4B_4} + 19\sqrt{B_4}\right) + |\lambda| L^2
\end{align}
and a new pair of constraints
\begin{align}
  A &= \left[\frac{1}{2\sqrt{B_4}} \left(B_2 + \frac{1}{2}B_3 + \frac{3}{4}B_4 -
        \frac{B_3^2}{4B_4}\right) + \frac{15}{2}\right] \nonumber \\
  &\quad {}\times \left[\frac{1}{2\sqrt{B_4}} \left(B_2 - \frac{3}{2}B_3 - \frac{5}{4}B_4 -
        \frac{B_3^2}{4B_4}\right) + \frac{13}{2}\right] \nonumber \\
  &\quad {}- \frac{1}{\sqrt{B_4}} \left(B_2 - \frac{1}{2}B_3 - \frac{1}{4}B_4 - \frac{B_3^2}{4B_4}
        \right) - 5 - \frac{L}{\sqrt{B_4}}(B_3+B_4), \\
  B_1 &= - \frac{B_3^2}{8B_4^2}(B_3-B_4) + \frac{1}{4B_4} \left[2B_2(B_3-B_4) - \frac{3}{2}B_3B_4
       - \frac{5}{2}B_4^2\right] \nonumber \\
  &\quad {}+ \frac{1}{\sqrt{B_4}}(3B_3-4B_4) - 2L\sqrt{B_4}.  
\end{align}
\par
%
%-----------------------------------------------------------------------------------------
%
The two pairs of constraints turn out to be compatible provided
\begin{equation}
  B_3 = B_4, \qquad B_2 = B_4 - 10\sqrt{B_4},
\end{equation}
from which we also obtain
\begin{equation}
\begin{split}
  &A = - B_4 - (2L+1)\sqrt{B_4} + \frac{35}{4}, \qquad B_1 = - B_4 - (2L+1)\sqrt{B_4}, \\
  &\frac{\eta}{|\lambda|} = \sqrt{B_4} - \frac{5}{2}, \qquad  \zeta = |\lambda|\sqrt{B_4}.
\end{split}
\end{equation}
\par
%
%-----------------------------------------------------------------------------------------
%
Hence, the potential
\begin{align}
  V(r) &= \frac{L(L+1)}{r^2} + |\lambda| \left(B_4 + (2L+1)\sqrt{B_4} - \frac{35}{4}\right) - |\lambda|
        \frac{B_4 + (2L+1)\sqrt{B_4} - \frac{35}{4}}{f^2} \nonumber \\
  &\quad {}- |\lambda| \frac{B_4 + (2L+1)\sqrt{B_4}}{f^4} + |\lambda| \frac{B_4 - 10\sqrt{B_4}}{f^6}
        + |\lambda| \frac{B_4}{f^8} + |\lambda| \frac{B_4}{f^{10}}, \label{eq:QES-potential-quater}
\end{align}
with corresponding superpotentials
\begin{equation}
\begin{split}
  &W(r) = - \frac{L+1}{r}f + |\lambda| \left(\sqrt{B_4} - \frac{5}{2}\right) \frac{r}{f} + |\lambda|
         \sqrt{B_4} \frac{r}{f^3} \left(1 + \frac{1}{f^2}\right), \\
  &W'(r) = - \frac{L+2}{r}f + |\lambda| \left(\sqrt{B_4} + \frac{5}{2}\right) \frac{r}{f} + |\lambda|
         \sqrt{B_4} \frac{r}{f^3} \left(1 + \frac{1}{f^2}\right),
\end{split}  \label{eq:W-W'-quater}
\end{equation}
has known first two levels with energies
\begin{equation}
\begin{split}
  E_0 &= |\lambda| \left[2B_4 - 2\sqrt{B_4} - \frac{13}{2} + L(4\sqrt{B_4} - 3) + L^2\right], \\
  E_1 &= E'_0 = |\lambda| \left[2B_4 + 10\sqrt{B_4} + \frac{23}{2} + L(4\sqrt{B_4} + 9) + L^2\right],
\end{split}
\end{equation}
and normalizable wavefunctions
\begin{align}
  \psi_0(r) &\propto r^{L+1} f^{\sqrt{B_4}-3} \exp\left(- \frac{1}{2}\sqrt{B_4}\frac{1}{f^2} 
        - \frac{1}{4}\sqrt{B_4}\frac{1}{f^4}\right), \\
  \psi_1(r) &\propto r^{L+1} f^{\sqrt{B_4}-3} [- (2L+3) + 3|\lambda|(2L+3+2\sqrt{B_4}) r^2
        \nonumber \\
  &\quad {}- 3|\lambda|^2(2L+3+2\sqrt{B_4}) r^4 + |\lambda|^3(2L+3+2\sqrt{B_4}) r^6]
        \nonumber \\
  &\quad \times \exp\left(- \frac{1}{2}\sqrt{B_4}\frac{1}{f^2} 
        - \frac{1}{4}\sqrt{B_4}\frac{1}{f^4}\right),
\end{align}
respectively. The single zero of $\psi_1(r)$ on $(0, 1/\sqrt{|\lambda|})$ is now located at $r_0 = \frac{1}{\sqrt{|\lambda|}} \{1 - [2\sqrt{B_4}/(2L+3+2\sqrt{B_4})]^{1/3}\}^{1/2}$.\par
%
%-----------------------------------------------------------------------------------------------------------
%
In Fig.~3, some examples of extended potentials (\ref{eq:QES-potential-ter}) and (\ref{eq:QES-potential-quater}) are plotted and the corresponding wavefunctions $\psi_0(r)$ and $\psi_1(r)$ of the former are displayed in Fig.~4.\par
%
%-----------------------------------------------------------------------------------------------------
%
\begin{figure}[h]
\begin{center}
\includegraphics{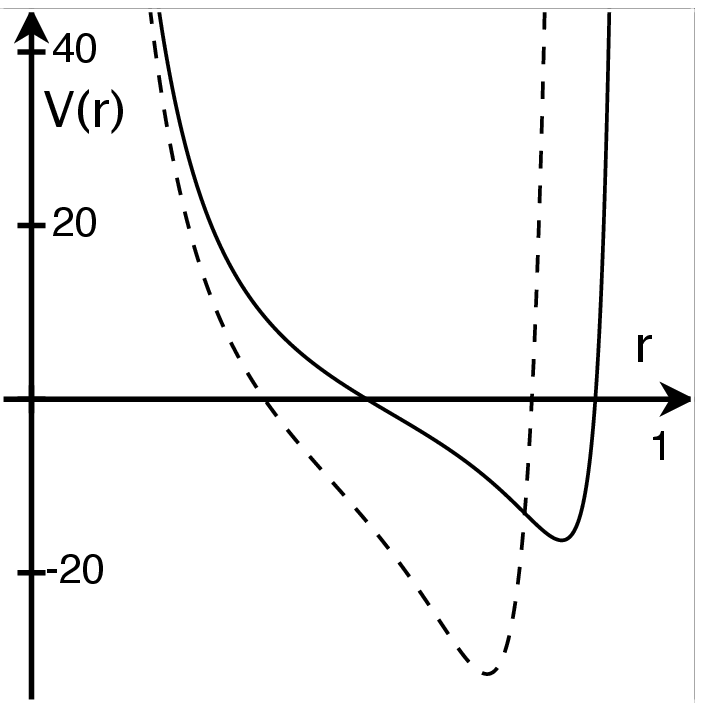}
\caption{Plots of potential (\ref{eq:QES-potential-ter}) with $-\lambda = L = B_2 = 1$ (solid line) and of potential (\ref{eq:QES-potential-quater}) with $-\lambda = L = B_4 = 1$ (dashed line). The ground and first excited state energies are $E_0 = 5/2$, $E_1 = 61/2$ for the former, and $E_0 = - 9/2$, $E_1 = 75/2$ for the latter.}
\end{center}
\end{figure}
\par
%
%-------------------------------------------------------------------------------------------------------------------------------
%
\begin{figure}[h]
\begin{center}
\includegraphics{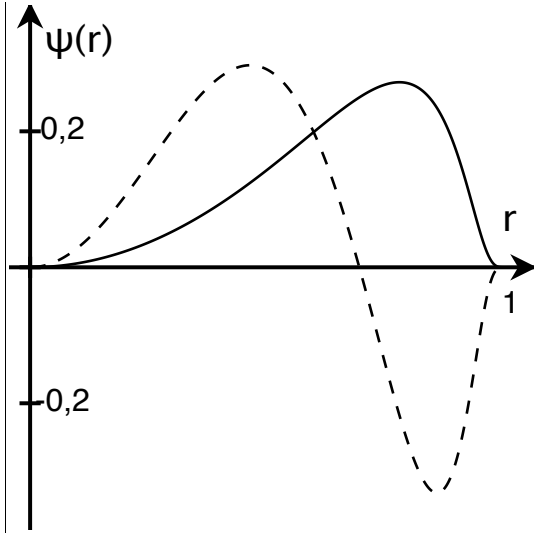}
\caption{Plots of ground state wavefunction $\psi_0(r)$ (solid line) and of first excited state wavefunction $\psi_1(r)$ (dashed line) for potential (\ref{eq:QES-potential-ter}) with $-\lambda = L = B_2 = 1$.}
\end{center}
\end{figure}
\par
%
%---------------------------------------------------------------------------------------------------------
%
It is obvious that we might continue applying the CDSI method to potentials of the first or second family with $m$ values higher than 2 in order to build potentials with known ground and first excited states. However, since the number of constraints would grow with $m$, the complexity of the method would increase correspondingly. Finding a general approach to build such potentials in DSUSY is therefore most desirable. This is the purpose of Section~IV.\par
%
%==========================================================
%
\section{GENERAL APPROACH FOR BUILDING POTENTIALS WITH TWO KNOWN EIGENSTATES IN DSUSY}

\setcounter{equation}{0}

Let us consider two pairs of DSUSY partner Hamiltonians $(\hat{H}_1, \hat{H}_2)$ and $(\hat{H}'_1, \hat{H}'_2)$, defined as in (\ref{eq:H-12}) and (\ref{eq:new-H-12}) in terms of two superpotentials $W(r)$ and $W'(r)$, respectively. From Eq.~(\ref{eq:new-rescaled-V}), it follows that the superpotentials are related by the equation
\begin{equation}
  W^2(r) + f(r) \frac{dW(r)}{dr} = W^{\prime2}(r) - f(r) \frac{dW'(r)}{dr} + E_1 - E_0,  \label{eq:W-W'-gen}
\end{equation}
where we have assumed that $E'_0 = E_1$. Let us now define the two functions
\begin{equation}
  W_+(r) = W'(r) + W(r), \qquad W_-(r) = W'(r) - W(r).  \label{eq:W_+-W_-}
\end{equation}
In terms of them, Eq.~(\ref{eq:W-W'-gen}) can be rewritten as
\begin{equation}
  f(r) \frac{dW_+(r)}{dr} = W_+(r) W_-(r) + E_1 - E_0,
\end{equation}
thus showing that $W_-(r)$ can be expressed in terms of $W_+(r)$ and of the energy difference $E_1 - E_0$ as
\begin{equation}
  W_-(r) = \frac{f(r) dW_+(r)/dr + E_0 - E_1}{W_+(r)}.  \label{eq:W_-}
\end{equation}
\par
%
%---------------------------------------------------------------------------------------------------
%
The general approach starts from two functions $W_+(r)$ and $W_-(r)$ that are compatible, i.e., such that Eq.~(\ref{eq:W_-}) is satisfied for some (yet unknown) positive constant $E_1-E_0$. Then the two superpotentials $W(r)$ and $W'(r)$ are determined from Eq.~(\ref{eq:W_+-W_-}). The starting potential $V_1(r)$ is obtained from Eq.~(\ref{eq:H-12}) and its first two wavefunctions $\psi_0(r)$ and $\psi_1(r)$ follow from Eqs.~(\ref{eq:gs-wf}) and (\ref{eq:es-wf}), respectively. On taking Eq.~(\ref{eq:A}) into account, the latter can be rewritten as
\begin{align}
  \psi_1(r) &\propto \left(- f\frac{d}{dr} - \frac{1}{2}\frac{df}{dr} + W\right) f^{-1/2} \exp\left(-
       \int^r \frac{W'(r')}{f(r')} dr'\right) \nonumber \\
  &\propto [W'(r) + W(r)]  f^{-1/2} \exp\left(-\int^r \frac{W'(r')}{f(r')} dr'\right) \nonumber \\
  &\propto W_+(r)  f^{-1/2} \exp\left(-\int^r \frac{W'(r')}{f(r')} dr'\right),
\end{align}
where, in the last step, definition (\ref{eq:W_+-W_-}) was used. The whole construction is of course valid provided the resulting functions $\psi_0(r)$ and $\psi_1(r)$ are normalizable on $(0, r_{\rm max})$.\par
%
%----------------------------------------------------------------------------------------------
% 
Our aim will now be to build QES potentials of type (\ref{eq:fam-1}) or (\ref{eq:fam-2}) with arbitrary $m \in 
\N^+$ and two known eigenstates by selecting the starting functions $W_{\pm}(r)$ appropriately. For such a purpose, we will be guided by the results obtained in Section~III.\par
%
%xxxxxxxxxxxxxxxxxxxxxxxxxxxxxxxxxxxxxxxxxxxxxxxxxxxxxxxxxxxxxxxxxxxxx
%
\subsection{First family of extended potentials}

{}For the first two members of the potential family (\ref{eq:fam-1}) (where $\lambda$ and $B_{2m}$ are assumed positive), we obtain from Eqs.~(\ref{eq:W-W'}) and (\ref{eq:W-W'-bis})
\begin{equation}
  W_+(r) = \frac{f}{r} [- (2L+3) + 2\lambda \sqrt{B_2}\, r^2], \qquad W_-(r) = \frac{r}{f} \left(- \frac{1}{r^2}
  + 2\lambda\right),
\end{equation}
if $m=1$, and
\begin{align}
  W_+(r) &= \frac{f}{r} [- (2L+3) + 2\lambda \sqrt{B_4}\, r^2 (1+f^2)] \nonumber \\
  & = \frac{f}{r} [- (2L+3) + 4\lambda \sqrt{B_4}\, r^2 + 2\lambda^2 \sqrt{B_4}\, r^4], \\
  W_-(r) &= \frac{r}{f} \left(- \frac{1}{r^2} + 4\lambda\right), \nonumber
\end{align}
if $m=2$.\par
%
%-------------------------------------------------------------------------------------------------
%
{}For any $m\in \N^+$, let us choose the functions
\begin{align}
  W_+(r) &= \frac{f}{r} \left[- (2L+3) + 2\lambda \sqrt{B_{2m}}\, r^2 \sum_{i=0}^{m-1} f^{2i}\right] 
       \nonumber \\
  &= \frac{f}{r} \left[- (2L+3) + 2 \sqrt{B_{2m}} \sum_{j=1}^m \binom{m}{j} (\lambda r^2)^j\right], 
        \label{eq:W_+-W_--1}\\
  W_-(r) &= \frac{r}{f} \left(- \frac{1}{r^2} + 2m\lambda\right). \nonumber
\end{align}
To be acceptable, this choice must be such that Eq.~(\ref{eq:W_-}) is satisfied for some positive constant $E_1-E_0$. It is straightforward to show that this is indeed so  for
\begin{equation}
  E_1 - E_0 = 2m\lambda (2L+3 + 2\sqrt{B_{2m}}).  \label{eq:constant}
\end{equation}
This result agrees with those that can be deduced from Eqs.~(\ref{eq:E0-E1}) and (\ref{eq:E0-E1-bis}) for $m=1$ and $m=2$, respectively.\par
%
%------------------------------------------------------------------------------------
%
{}From Eq.~(\ref{eq:W_+-W_--1}), we obtain for the two superpotentials
\begin{align}
  W(r) &= - \frac{L+1}{r}f - \left(m+\frac{1}{2}\right)\frac{\lambda r}{f} + \lambda \sqrt{B_{2m}}\, r 
       \sum_{i=0}^{m-1} f^{2i+1}, \label{eq:W-1} \\
  W'(r) &= - \frac{L+2}{r}f + \left(m+\frac{1}{2}\right)\frac{\lambda r}{f} + \lambda \sqrt{B_{2m}}\, r
       \sum_{i=0}^{m-1} f^{2i+1}.  
\end{align}
\par
%
%-------------------------------------------------------------------------------------------------
% 
After a rather lengthy, but straightforward calculation, Eqs.~(\ref{eq:H-12}) and (\ref{eq:W-1}) yield
\begin{align}
  V_1(r) &= \frac{L(L+1)}{r^2} + \lambda\left[\left(L+m+\frac{3}{2}\right)^2 + (2m+2)\sqrt{B_{2m}}
       \right] - \frac{\lambda \left(m+\frac{1}{2}\right) \left(m-\frac{1}{2}\right)}{f^2} \nonumber \\
  &\quad {}- \lambda [B_{2m} + (2L+1) \sqrt{B_{2m}}] \sum_{i=1}^{m-1} f^{2i} - \lambda [B_{2m} +
       (2L+4m+3) \sqrt{B_{2m}}] f^{2m} \nonumber \\
  &\quad {}+ \lambda B_{2m} \sum_{i=m+1}^{2m} f^{2i}.  \label{eq:rescaled-V-fam-1}
\end{align}
This corresponds to potential $V(r)$ of Eq.~(\ref{eq:fam-1}) with
\begin{equation}
\begin{split}
  &A = \left(m+\frac{1}{2}\right)\left(m-\frac{1}{2}\right), \\
  &B_1 = B_2 = \cdots = B_{m-1} = - B_{2m} - (2L+1) \sqrt{B_{2m}}, \\
  &B_m = - B_{2m} - (2L+4m+3)\sqrt{B_{2m}}, \\
  &B_{m+1} = B_{m+2} = \cdots = B_{2m},
\end{split}  \label{eq:param-1}
\end{equation}
and to a ground state energy given by
\begin{equation}
  \lambda A - E_0 = \lambda \left[\left(L+m+\frac{3}{2}\right)^2 + (2m+2) \sqrt{B_{2m}}\right]
\end{equation}
or 
\begin{equation}
  E_0 = - \lambda \left[(2m+2) \sqrt{B_{2m}} + 3m + \frac{5}{2} + (2m+3)L + L^2\right].
  \label{eq:E0}
\end{equation}
Inserting (\ref{eq:E0}) in (\ref{eq:constant}), we get
\begin{equation}
  E_1 = \lambda \left[(2m-2)\sqrt{B_{2m}} + 3m - \frac{5}{2} + (2m-3)L - L^2\right]
\end{equation}
for the first excited state energy of potential (\ref{eq:fam-1}) with parameters (\ref{eq:param-1}).\par
%
%---------------------------------------------------------------------------------------------------
%
{}From (\ref{eq:rescaled-V-fam-1}), we can also determine the partner $V_2(r) = V_1(r) + 2f(r) dW/dr$ of $V_1(r)$. The result reads
\begin{equation}
  V_2(r) = \frac{L'(L'+1)}{r^2} + \lambda A' - \frac{\lambda A'}{f^2} + \lambda \sum_{k=1}^{2m}
  B'_k f^{2k} + R
\end{equation}
with
\begin{equation}
\begin{split}
  &L' = L+1, \\
  &A' = \left(m+\frac{3}{2}\right)\left(m+\frac{1}{2}\right), \\
  &B'_1 = B'_2 = \cdots = B'_m = - B_{2m} - (2L+3) \sqrt{B_{2m}}, \\
  &B'_{m+1} = B'_{m+2} = \cdots = B'_{2m} = B_{2m}, \\
  &R = \lambda \left[2m\sqrt{B_{2m}} + m + \frac{3}{2} + (2m+3)L + L^2\right].
\end{split}
\end{equation}
\par
%
%--------------------------------------------------------------------------------------------
% 
{}Finally, the first two (normalizable) wavefunctions are given by 
\begin{equation}
  \psi_0(r) \propto r^{L+1} f^m \exp\left[- \frac{1}{2} \sqrt{B_{2m}} \sum_{j=1}^m \frac{1}{j}
  \binom{m}{j} (\lambda r^2)^j\right]
\end{equation}
and
\begin{align}
  \psi_1(r) &\propto r^{L+1} f^{-m} \left[- (2L+3) + 2\sqrt{B_{2m}} \sum_{j=1}^m \binom{m}{j}
       (\lambda r^2)^j \right] \nonumber \\
  &\quad {}\times \exp\left[- \frac{1}{2} \sqrt{B_{2m}} \sum_{j=1}^m \frac{1}{j} \binom{m}{j} 
       (\lambda r^2)^j\right], 
\end{align}
respectively. Observe that $\psi_1(r)$ has a single zero  on $(0,\infty)$ located at $r_0 = \frac{1}{\sqrt{\lambda}} \{[(2L+3 + 2\sqrt{B_{2m}})/(2\sqrt{B_{2m}})]^{1/m} - 1\}^{1/2}$.\par
%
%xxxxxxxxxxxxxxxxxxxxxxxxxxxxxxxxxxxxxxxxxxxxxxxxxxxxxxxxxxxxxxxxxxxxxx
%
\subsection{Second family of extended potentials}

{}From Eqs.~(\ref{eq:W-W'-ter}) and (\ref{eq:W-W'-quater}) valid for the first two members of family (\ref{eq:fam-2}) (with $\lambda<0$ and $B_{2m}>0$), we get
\begin{align}
  W_+(r) &= \frac{f}{r} \left[- (2L+3) + 2|\lambda|\sqrt{B_2}\, r^2 \left(\frac{1}{f^2} + \frac{1}{f^4}
       \right)\right] \nonumber \\
  &= \frac{1}{rf^3} [- (2L+3) + 2|\lambda|(2L+3 + 2\sqrt{B_2})r^2 - |\lambda|^2 (2L+3 + 2\sqrt{B_2})
       r^4], \\
  W_-(r) &= \frac{r}{f} \left(- \frac{1}{r^2} + 4|\lambda|\right),
\end{align}
and
\begin{align}
  W_+(r) &= \frac{f}{r} \left[- (2L+3) + 2|\lambda|\sqrt{B_4}\, r^2 \left(\frac{1}{f^2} + \frac{1}{f^4}
       + \frac{1}{f^6}\right)\right] \nonumber \\
  &= \frac{1}{rf^5} [- (2L+3) + 3|\lambda|(2L+3 + 2\sqrt{B_4})r^2 - 3|\lambda|^2 (2L+3 + 2\sqrt{B_4})
       r^4 \nonumber \\
  &\quad {}+ |\lambda|^3 (2L+3 + 2\sqrt{B_{4}}) r^6], \\
  W_-(r) &= \frac{r}{f} \left(- \frac{1}{r^2} + 6|\lambda|\right),
\end{align}
respectively.\par
%
%--------------------------------------------------------------------------------------------------------
%
{}For any $m\in\N^+$, let us choose the functions
\begin{align}
  W_+(r) &= \frac{f}{r} \left[- (2L+3) + 2|\lambda| \sqrt{B_{2m}}\, r^2 \sum_{i=0}^m \frac{1}{f^{2i+2}}
       \right] \nonumber \\
  &= \frac{1}{rf^{2m+1}} \left[- (2L+3) - (2L+3+2\sqrt{B_{2m}}) \sum_{j=1}^{m+1} (-1)^j 
       \binom{m+1}{j} (|\lambda|r^2)^j\right], \\
  W_-(r) &= \frac{r}{f} \left[- \frac{1}{r^2} + (2m+2) |\lambda]\right]. \nonumber 
\end{align}
This choice is acceptable because Eq.~(\ref{eq:W_-}) is satisfied with
\begin{equation}
  E_1 - E_0 = (2m+2) |\lambda| (2L+3 + 2\sqrt{B_{2m}}).  \label{eq:constant-bis}
\end{equation}
\par
%
%----------------------------------------------------------------------------------------------
%
The two superpotentials $W(r)$ and $W'(r)$ are given by
\begin{equation}
  W(r) = - \frac{L+1}{r}f + \frac{r}{f} |\lambda| \left(\sqrt{B_{2m}} - \frac{2m+1}{2}\right) + |\lambda|
  \sqrt{B_{2m}}\, r \sum_{i=1}^m \frac{1}{f^{2i+1}}  \label{eq:W-2}
\end{equation}
and
\begin{equation}
  W'(r) = - \frac{L+2}{r}f + \frac{r}{f} |\lambda| \left(\sqrt{B_{2m}} + \frac{2m+1}{2}\right) + |\lambda|
  \sqrt{B_{2m}}\, r \sum_{i=1}^m \frac{1}{f^{2i+1}},
\end{equation}
respectively.\par
%
%-----------------------------------------------------------------------------------------------
%
Then Eqs.~(\ref{eq:H-12}) and (\ref{eq:W-2}) yield
\begin{align}
  V_1(r) &= \frac{L(L+1)}{r^2} - |\lambda| \left(L+1+\sqrt{B_{2m}} - \frac{2m+1}{2}\right)^2
       \nonumber \\ 
  &\quad {}+ \frac{|\lambda|}{f^2} \left[- B_{2m} - (2L+1) \sqrt{B_{2m}} + \frac{1}{4}(2m+1)
       (2m+3)\right] \nonumber \\
  &\quad {} - |\lambda| [B_{2m} + (2L+1) \sqrt{B_{2m}}] \sum_{i=2}^m \frac{1}{f^{2i}} + \frac{|\lambda|}
       {f^{2m+2}} [B_{2m} - 2(2m+1) \sqrt{B_{2m}}] \nonumber \\
  &\quad {} + |\lambda| B_{2m} \sum_{i=m+2}^{2m+1} \frac{1}{f^{2i}}.
\end{align}
The corresponding starting potential is therefore $V(r)$ of Eq.~(\ref{eq:fam-2}) with
\begin{equation}
\begin{split}
  &A = - B_{2m} - (2L+1) \sqrt{B_{2m}} + \frac{1}{4}(2m+1)(2m+3), \\
  &B_1 = B_2 = \cdots = B_{m-1} = - B_{2m} - (2L+1) \sqrt{B_{2m}}, \\
  &B_m = B_{2m} - 2(2m+1) \sqrt{B_{2m}}, \\
  &B_{m+1} = B_{m+2} = \cdots = B_{2m},
\end{split}
\end{equation}
and its ground state energy is given by
\begin{equation}
  - |\lambda|A - E_0 = - |\lambda| \left(L+1+\sqrt{B_{2m}} - \frac{2m+1}{2}\right)^2
\end{equation}
or
\begin{equation}
  E_0 = |\lambda| \left[2B_{2m} - 2(m-1)\sqrt{B_{2m}} - 3m - \frac{1}{2} + L (4\sqrt{B_{2m}} - 2m
  + 1) + L^2\right].  \label{eq:E0-bis}
\end{equation}
Combining Eqs.~(\ref{eq:constant-bis}) and (\ref{eq:E0-bis}) yields the first excited state energy
\begin{equation}
  E_1 = |\lambda| \left[2B_{2m} + 2(m+3)\sqrt{B_{2m}} + 3m + \frac{11}{2} + L (4\sqrt{B_{2m}}
  + 2m +5) + L^2\right].
\end{equation}
\par
%
%-----------------------------------------------------------------------------------------------
%
{}For the partner $V_2(r) = V_1(r) + 2f(r) dW/dr$ of $V_1(r)$, we obtain 
\begin{equation}
  V_2(r) = \frac{L'(L'+1)}{r^2} - |\lambda|A' + \frac{|\lambda|A'}{f^2} + |\lambda| \sum_{k=1}^{2m}
  \frac{B'_k}{f^{2k+2}} + R
\end{equation}
with
\begin{equation}
\begin{split}
  &L' = L + 1, \\
  &A' = - B_{2m} - (2L+3)\sqrt{B_{2m}} + \frac{1}{4}(2m+1)(2m-1), \\
  &B'_1 = B'_2 = \cdots = B'_{m-1} = - B_{2m} - (2L+3)\sqrt{B_{2m}}, \\
  &B'_m = B'_{m+1} = \cdots = B'_{2m} = B_{2m}, \\
  &R = |\lambda| \left[- 2B_{2m} + 2(m-2)\sqrt{B_{2m}} + m - \frac{1}{2} + L(-4\sqrt{B_{2m}} + 2m
        -1) - L^2\right].
\end{split}
\end{equation}
\par
%
%----------------------------------------------------------------------------------------------
%
{}Finally, the ground state and first excited state wavefunctions of potential $V(r)$ are given by
\begin{equation}
  \psi_0(r) \propto r^{L+1} f^{\sqrt{B_{2m}} - m - 1} \exp\left(- \frac{1}{2} \sqrt{B_{2m}} \sum_{i=1}^m
  \frac{1}{if^{2i}}\right)
\end{equation}
and
\begin{align}
  \psi_1(r) &\propto r^{L+1} f^{\sqrt{B_{2m}} - m - 1} \left[- (2L+3) - (2L+3+2\sqrt{B_{2m}})
       \sum_{j=1}^{m+1} (-1)^j \binom{m+1}{j} (|\lambda| r^2)^j\right] \nonumber \\
  &\quad {}\times \exp\left(- \frac{1}{2} \sqrt{B_{2m}} \sum_{i=1}^m \frac{1}{if^{2i}}\right), 
\end{align}
respectively. The zero of $\psi_1(r)$ in the interval $(0, 1/\sqrt{|\lambda|})$ is located at $r_0 = \frac{1}{\sqrt{|\lambda|}} \{1 - [2\sqrt{B_{2m}}/(2L+3+2\sqrt{B_{2m}})]^{1/(m+1)}\}^{1/2}$.\par
%
%=========================================================
%
\section{CONCLUSION}

In the present paper, we have reconsidered the two families of extensions of the oscillator in a $d$-dimensional constant-curvature space that we had recently introduced and for which some bound state solutions had been obtained for some ad hoc values  of their parameters by using the functional Bethe ansatz method. Here, we have shown that other approaches based on DSUSY enable us to determine potentials with known ground and first excited eigenstates.\par
%
%----------------------------------------------------------------------------------------------
%
In the first method, we have extended the DSI symmetry, valid for the oscillator alone, by completing it with some constraints relating the potential parameters, thereby getting potentials that are CDSI. Considering the next step in the construction of a partner potential hierarchy, we have determined another set of constraints among the potential parameters. Solving the compatibility conditions between the two sets of constraints has made it possible to algebraically generate QES extensions of the oscillator with two known eigenstates (the ground and first excited ones). Explicit calculations have been carried out for the first two members of the two extension families.\par
%
%------------------------------------------------------------------------------------------------
% 
In order to get explicit outcomes valid for arbitrary members of the two families, we have then devised a general method, wherein the first two DSUSY superpotentials (and hence the potentials and their first two eigenstates) are expressed in terms of a function $W_+(r)$ (and its accompanying function $W_-(r)$). From the results obtained for $W_{\pm}(r)$ for the first two members of the families, we have proposed some general formulas for $W_{\pm}(r)$, which have enabled us to solve the problem in full generality.\par
%
%----------------------------------------------------------------------------------------------
%
Applying the methods developed here to other types of radial potential, such as the Kepler-Coulomb one, and/or other types of deformation function $f(r)$ would be interesting topics for future investigation.\par
%
%=================================================================
%


\newpage

\begin{thebibliography}{99}

\bibitem{mathews}
P.\ M.\ Mathews and M.\ Lakshmanan,
Q.\ Appl.\ Math.\ {\bf 32}, 215 (1974).

\bibitem{carinena04a}
J.\ F.\ Cari\~nena, M.\ F.\ Ra\~nada, M.\ Santander, and M.\ Senthilvelan, 
Nonlinearity {\bf 17}, 1941 (2004).

\bibitem{carinena04b}
J.\ F.\ Cari\~nena, M.\ F.\ Ra\~nada, and M.\ Santander,
Rep.\ Math.\ Phys.\ {\bf 54}, 285 (2004).

\bibitem{carinena07a}
J.\ F.\ Cari\~nena, M.\ F.\ Ra\~nada, and M.\ Santander,
Ann.\ Phys.\ {\bf 322}, 434 (2007).

\bibitem{schulze}
A.\ Schulze-Halberg and J.\ R.\ Morris,
J.\ Phys.\ A: Math.\ Theor.\ {\bf 45}, 305301 (2012).

\bibitem{carinena07b}
J.\ F.\ Cari\~nena, M.\ F.\ Ra\~nada, and M.\ Santander,
Ann.\ Phys.\ {\bf 322}, 2249 (2007).

\bibitem{carinena07c}
J.\ F.\ Cari\~nena, M.\ F.\ Ra\~nada, and M.\ Santander,
J.\ Math.\ Phys.\ {\bf 48}, 102106 (2007).

\bibitem{cq15}
C.\ Quesne,
Phys.\ Lett.\ A {\bf 379}, 1589 (2015).

\bibitem{carinena12}
J.\ F.\ Cari\~nena, M.\ F.\ Ra\~nada, and M.\ Santander,
J.\ Phys.\ A: Math.\ Theor.\ {\bf 45}, 265303 (2012).

\bibitem{cq16}
C.\ Quesne,
J.\ Math.\ Phys.\ {\bf 57}, 102101 (2016).

\bibitem{gomez}
D.\ G\'omez-Ullate, Y.\ Grandati, and R.\ Milson,
J.\ Math.\ Phys.\ {\bf 55}, 043510 (2014).

\bibitem{cq04}
C.\ Quesne and V.\ M.\ Tkachuk,
J.\ Phys.\ A: Math.\ Gen.\ {\bf 37}, 4267 (2004).

\bibitem{bagchi}
B.\ Bagchi, A.\ Banerjee, C.\ Quesne, and V.\ M.\ Tkachuk,
J.\ Phys.\ A: Math.\ Gen.\ {\bf 38}, 2929 (2005).

\bibitem{cq17}
C.\ Quesne,
J.\ Math.\ Phys.\ {\bf 58}, 052104 (2017).

\bibitem{turbiner87}
A.\ V.\ Turbiner and A.\ G.\ Ushveridze,
Phys.\ Lett.\ A {\bf 126}, 181 (1987).

\bibitem{turbiner88}
A.\ V.\ Turbiner,
Commun.\ Math.\ Phys.\ {\bf 118}, 467 (1988).

\bibitem{ushveridze}
A.\ G.\ Ushveridze,
{\sl Quasi-Exactly Solvable Models in Quantum Mechanics} (IOP, Bristol, 1994).

\bibitem{gonzalez}
A.\ Gonz\'alez-L\'opez, N.\ Kamran, and P.\ J.\ Olver,
Commun.\ Math.\ Phys.\ {\bf 153}, 117 (1993).

\bibitem{turbiner16}
A.\ V.\ Turbiner,
Phys.\ Rep.\ {\bf 642}, 1 (2016).

\bibitem{ronveaux}
A.\ Ronveaux,
{\sl Heun Differential Equations} (Oxford University Press, Oxford, 1995).

\bibitem{gaudin}
M.\ Gaudin,
{\sl La Fonction d'Onde de Bethe} (Masson, Paris, 1983).

\bibitem{ho}
C.-L.\ Ho,
Ann.\ Phys.\ {\bf 323}, 2241 (2008).

\bibitem{zhang12}
Y.-Z.\ Zhang,
J.\ Phys.\ A: Math.\ Theor.\ {\bf 45}, 065206 (2012).

\bibitem{gendenshtein}
I.\ E.\ Gendenshtein,
JETP Lett.\ {\bf 38}, 356 (1983).

\bibitem{cooper}
F.\ Cooper, A.\ Khare, and U.\ Sukhatme,
Phys.\ Rep.\ {\bf 251}, 267 (1995).

\bibitem{gango}
A.\ Gangopadhyaya, A.\ Khare, and U.\ P.\ Sukhatme,
Phys.\ Lett.\ A {\bf 208}, 261 (1995).

\bibitem{jatkar}
D.\ P.\ Jatkar, C.\ Nagaraja Kumar, and A.\ Khare,
Phys.\ Lett.\ A {\bf 142}, 200 (1989).

\bibitem{roy}
P.\ Roy and Y.\ P.\ Varshni,
Mod.\ Phys.\ Lett.\ A {\bf 6}, 1257 (1991).

\bibitem{chakrabarti}
B.\ Chakrabarti,
J.\ Phys.\ A: Math.\ Theor.\ {\bf 41}, 405301 (2008).

\bibitem{bera}
S.\ Bera, B.\ Chakrabarti, and T.\ K.\ Das,
Phys.\ Lett.\ A {\bf 381}, 1356 (2017).

\bibitem{tkachuk98}
V.\ M.\ Tkachuk,
Phys.\ Lett.\ A {\bf 245}, 177 (1998).

\bibitem{tkachuk01}
V.\ M.\ Tkachuk,
J.\ Phys.\ A: Math.\ Gen.\ {\bf 34}, 6339 (2001).

\bibitem{kuliy}
T.\ V.\ Kuliy and V.\ M.\ Tkachuk,
J.\ Phys.\ A: Math.\ Gen.\ {\bf 32}, 2157 (1999).

\bibitem{mustafa}
O.\ Mustafa and S.\ H.\ Mazharimousavi,
Int.\ J.\ Theor.\ Phys.\ {\bf 46}, 1786 (2007).

\end{thebibliography}
\end{document}